\documentstyle[aps,psfig]{revtex} 
\newcommand{\be}{\begin{equation}} 
\newcommand{\ee}{\end{equation}} 
\newcommand{\ben}{\begin{eqnarray}} 
\newcommand{\een}{\end{eqnarray}} 
\begin{document} 
 
\twocolumn[\hsize\textwidth\columnwidth\hsize\csname 
@twocolumnfalse\endcsname 
 
\title{Entrapment of a Network of Domain Walls} 
\author{D. Bazeia and F. A. Brito} 
\address{Departamento de F\'\i sica, Universidade Federal da Para\'\i ba,\\ 
Caixa Postal 5008, 58051-970 Jo\~ao Pessoa, Para\'\i ba, Brazil} 
\date{\today} 
 
\maketitle 

\begin{abstract}
We explore the idea of a network of defects to live inside a domain wall in
models of three real scalar fields, engendering the $Z_2\times Z_3$ symmetry.
The field that governs the $Z_2$ symmetry generates a domain wall, and
entraps the hexagonal network formed by the three-junctions of the
model of two scalar fields that describes the remaining $Z_3$ symmetry.
If the host domain wall bends to the spherical form, in the thin wall
approximation there may appear non-topological structures hosting networks
that accept diverse patterns. If $Z_3$ is also broken,
the model may generate a buckyball containing sixty junctions,
a fullerene-like structure. Applications to cosmology are outlined.
\end{abstract}
\vskip1pc]
  
Domain walls appear in diverse branches of physics, as for instance in
systems of condensed matter that present ferromagnetic \cite{esc81},
ferroelectric \cite{sle98} and other properties \cite{wal97}, and also
in cosmology \cite{ktu90,vsh96}. They arise in systems with at least two
isolated degenerate minima, and in field theory they usually live in three
spatial dimensions as bidimensional objects, seen as immersions into $(3,1)$
dimensions of static solutions of $(1,1)$ dimensional models that engender
the $Z_2$ symmetry. The standard domain wall presents no internal structure,
but there are models where they may entrap field configurations that engender
non-trivial behavior. This idea follows as in
Refs.~{\cite{wit85,lsh85,mac88}}, and in the more recent Refs.~{\cite{mor,ft}.
Other investigations include for instance supersymmetry \cite{susy1},
supergravity \cite{cso97}, and the recent applications to polymers {\cite{cm}}.

Although domain walls may be dangerous \cite{ktu90,vsh96} to cosmological
applications, they have found their way into cosmology as for instance seeds
for the formation of non-topological structures. This possibility appears
in Refs.~{\cite{lee87,fgg88,mca95,mba96}}, where the discrete
symmetry is changed to an approximate symmetry, or in Ref.~{\cite{clo96}},
with the discrete symmetry biased so that domains of distinct but
degenerate vacua spring unequally. The non-topological structures
may be stable, but now stability requires the presence of conserved charges,
of bosonic and/or fermionic origin.

Domain walls may also be of interest when they host non-trivial structures.
We illustrate this point using the model introduced in the first work of
Ref.~{\cite{ft}}, describing the pair of fields $(\phi,\chi)$ via the
superpotential $W(\phi,\chi)=-\phi+(1/3)\phi^3+r\,\phi\,\chi^2$.
Here $r$ is a parameter, real and dimensionless, that couples the two fields.
The system is described by a quartic potential, and we use natural units,
working with dimensionless space and time variables, and fields. The equations
of motion for field configurations $\phi=\phi(z)$ and $\chi=\chi(z)$ are
solved by solutions of the first order differential equations
${d\phi/dz}=-1+\phi^2+r\chi^2$ and ${d\chi/dz}=2r\phi\chi$.
In this model, the sector connecting the minima $(\pm1,0)$ is a BPS sector,
with energy density or tension $t=4/3$. For $r>0$ this BPS sector admits two
different types of static solutions: the one-field solutions
$\phi_1(z)=\tanh(z)$ and $\chi_1=0$, and the two-field solutions
$\phi_2(z)=-\tanh(2\,rz)$ and $\chi_2(z)=a(r)/\cosh(2\,rz)$,
with $a^2(r)=1/r-2$, valid for $0<r<1/2$.
We are working in $(3,1)$ space-time dimensions, so the one-field solution
represents a standard domain wall, while the two-field solution appears as
a domain wall having internal structure. As $z$ varies in $(-\infty,\infty)$,
in configuration space the vectors $(\phi_k\, ,\chi_k)$, $k=1,2$,
describe a straight line segment $(k=1)$ and an elliptic arc $(k=2)$,
resembling light in the linearly and elliptically polarized cases,
respectively. These solutions also appear in condensed matter, and there
they are named Ising and Bloch walls, respectively \cite{esc81,sle98,wal97}.
They appear as solutions of the anisotropic $XY$ system, a model used to
describe ferromagnetic transition in magnetic systems. The Bloch walls
can be seen as chiral interfaces, and may be used to describe more complex
phenomena, as for instance in the applications where chirality is also
broken \cite{nice}.

In Ref.~{\cite{bbr00}} the idea of nesting a network of defects inside a
domain wall has been presented. This possibility may
appear in models with three real scalar fields, engendering the
$Z_2\times Z_3$ symmetry. In the present work we offer a model that contains
the basic mechanisms behind this idea. The model will ultimately lead to the
scenario of a domain wall hosting a network of defects, which may have direct
interest to physics, as we show when we explore the pattern of the nested
network in the case we allow the underlying $Z_2\times Z_3$ symmetry
to be broken.

We first develop the idea of a domain wall hosting a network of defects,
in a model described by three real scalar fields, with the (dimensionless)
potential,
\ben
\label{p}
V(\sigma,\phi,\chi)&=&\frac{2}{3}\,\left(\sigma^2-\frac{9}{4}\right)^2+
\left(r\,\sigma^2-\frac{9}{4}\right)\,(\phi^2+\chi^2)
\nonumber\\
& &+(\phi^2+\chi^2)^2-\phi\,(\phi^2-3\,\chi^2)
\een
Here $r$ couples $\sigma$ to the pair of fields $(\phi,\chi)$. This potential
is polynomial, and contains up to the fourth order power in the fields.
Thus, it behaves standardly in $(3,1)$ space-time dimensions. Also, it
presents discrete $Z_2\times Z_3$ symmetry. We set $(\phi,\chi)\to(0,0)$,
to get the projection $V(\sigma,0,0)\to V(\sigma)=(2/3)\,(\sigma^2-9/4)^2$.
The projected potential presents $Z_2$ symmetry, and can be written with
the superpotential $W(\sigma)=(2\sqrt{3}/9)\sigma^3-(3\sqrt{3}/2)\sigma$,
in the form $V=(1/2)(dW/d\sigma)^2$. The reduced model supports the explicit
configurations $\sigma_h(z)=\pm\,(3/2)\,\tanh(\sqrt{3} z)$. The tension
of the host wall is $t_h=3\,\sqrt{3}=(3/2)\,m_h$, where $m_h$ represents
the mass of the elementary $\sigma$ meson. Also, the width of the wall
is such that $l_h\sim1/\sqrt{3}$.

The potentials projected inside $(\sigma\to0)$ and outside
$(\sigma\to\pm\,3/2)$ the host domain wall are $V_{in}(\phi,\chi)$
and $V_{out}(\phi,\chi)$. Inside the wall we have
\ben
V_{in}(\phi,\chi)&=&(\phi^2+\chi^2)^2-\phi\,(\phi^2-3\,\chi^2)\nonumber
\\
& &-\frac{9}{4}(\phi^2+\chi^2)+\frac{27}{8}
\een
This potential engenders the $Z_3$ symmetry, and there are three
global minima, at the points $v^{in}_1=(3/2)(1,0)$ and
$v^{in}_{2,3}=(3/4)(-1,\pm\sqrt{3})$, which define an
equilateral triangle. Outside the wall we get
\ben
V_{out}(\phi,\chi)&=&(\phi^2+\chi^2)^2-\phi\,
(\phi^2-3\,\chi^2)\nonumber\\
& &+\frac{9}{4}\,(r-1)\,(\phi^2+\chi^2)
\een
$V_{out}$ also engenders the $Z_3$ symmetry, but now the minima depend
on $r$. We can adjust $r$ such that $r>9/8$, which is the condition
for the fields $\phi$ and $\chi$ to develop no non-zero vacuum expectation
value outside the host domain wall, ensuring that the model supports no
domain defect outside the host domain wall. The restriction of considering
quartic potentials forbids the possibility of describing the $Z_3$ portion
of the model with the complex superpotential used in {\cite{susy2}; see
also Ref.~{\cite{more}}.

We investigate the masses of the elementary $\phi$ and $\chi$ mesons.
Inside the wall they degenerate to the single value $m_{in}=3\sqrt{3/2}$.
Outside the wall, for $r>9/8$ they also degenerate to a single value,
$m_{out}(r)=3\sqrt{(r-1)/2}$, which depends on $r$. We see that
$m_{out}(r=4)=m_{in}$. Also, $m_{out}(r)> m_{in}$ for $r>4$,
and $m_{out}(r)< m_{in}$ for $r$ in the interval $(9/8,4)$.

We study linear stability of the classical solutions
$\sigma=\sigma_h(z)$ and $(\phi,\chi)=(0,0)$. The fields $\phi$ and
$\chi$ vanish classically, and their fluctuations $(\eta_n,\xi_n)$ decouple.
The procedure leads to two equations for the fluctuations,
that degenerate to the single Schr\"odinger-like equation
\be
-\,\frac{d^2\psi_n(z)}{dz^2}+\frac{9}{2}\,V(z)\;
\psi_n(z)=\,w_n^2\,\psi_n(z)
\ee
Here $V(z)=-1+r\,\tanh^2{\sqrt{3}z}$.
This equation is of the modified P\"oschl-Teller type, and can be
examined analytically. The lowest eigenvalue is
$w_0^2=(3/2)\sqrt{6\,r+1\,}-6$. There is instability for $r<5/2$, showing
that the host domain wall with $(\phi,\chi)=(0,0)$ is unstable and therefore
relax to lower energy configurations, with $(\phi,\chi)\neq(0,0)$ for
$r<5/2$. Inside the host domain wall the sigma field vanishes, and the model
is governed by the potential $V_{in}(\phi,\chi)$, which consequently may
allow the presence of non-trivial $(\phi,\chi)$ configurations. The host
domain wall entraps the system described by $V_{in}(\phi,\chi)$ for the
parameter $r$ in the interval $(9/8,5/2)$. In this interval we have
$m_{out}<m_{in}$, showing that it is not energetically favorable for the
elementary $\phi$ and $\chi$ mesons to live inside the wall for
$r\in(9/8,5/2)$. The model automatically suppress backreactions of the
$\phi$ and $\chi$ mesons into the defects that may appear inside
the host domain wall.

In Ref.~{\cite{bbr00}} the potential inside the wall was shown to admit
a network of domain walls, in the form of a hexagonal array of domain walls.
In the thin wall approximation the network may be
represented by the solutions
\ben 
\phi_1&=&\frac{3}{8}+\frac{9}{8} 
\tanh\left(\frac{1}{2}\sqrt{\frac{27}{8}}\,(y+\sqrt{3}x)\right) 
\\ 
\chi_1&=&\frac{3}{8}\sqrt{3}-\frac{3}{8}\sqrt{3} 
\tanh\left(\frac{1}{2}\sqrt{\frac{27}{8}}\,(y+\sqrt{3}x)\right) 
\een 
and by $(\phi_k,\chi_k)$, obtained by rotating the pair
$(\phi_1,\chi_1)$ by $2(k-1)\pi/3$, for $k=2,3$. We identify the
space $(\phi,\chi)$ with $(x,y)$, so rotations in $(\phi,\chi)$ also
rotates the plane $(x,y)$ accordingly. The energy or tension of the
individual defects in the network is given by, in the thin wall
approximation $t_{n}=(27/8)\sqrt{3/2}=(9/8)\,m_{in}$.
In the nested network, the width of each defect obeys $l_n\sim\sqrt{8/27}$.
This shows that $l_h/l_n=3/2\sqrt{2}$, and so the host domain wall is
slightly thicker than the defects in the nested network. In the thin
wall approximation, the potential $V_{in}(\phi,\chi)$ allows the formation
of three-junctions as reactions that occur exothermically, and the nested
array of thin wall configurations is stable. In Fig.~1 we depict the
hexagonal network of defects inside the domain wall, in the thin wall
approximation. The dashed lines show equilateral triangles, that belong
to the dual lattice. Both the hexagonal network and the dual triagular
network are composed of equilateral polygons, a fact that follows in
accordance with the $Z_3$ symmetry.

\begin{figure}
\centerline{\psfig{figure=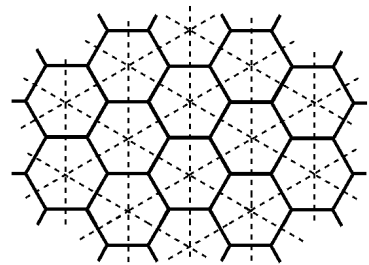,height=3.5cm}} 
\vspace{0.5cm}
\caption{The equilateral hexagonal network of defects, that may live
inside the host domain wall. The dashed lines show the dual lattice,
formed by equilateral triangles.}
\end{figure}

We now explore the breaking of the $Z_2\times Z_3$ symmetry
of the model. The simplest case refers to the breaking of the $Z_3$ symmetry,
without breaking the remaining $Z_2$ symmetry. We consider the case
of breaking the internal $Z_3$ symmetry in the following way. We take
for instance the vacuum state $v_1^{in}=(3/2)\,(1,0)$, and change
its position to a location farther from or closer to the other minima of
the system, increasing or decreasing the angle between two of the three
defects. We illustrate this situation in Fig.~2. We do this
without removing the degeneracy of the three minima. This mechanism
changes $Z_3\to Z_2$, so we refer to it as the minimal breaking. 

We notice that the energy of the defect depends on the distance between the
two minima the defect connects, and goes with the cube of it. If the $Z_3$
symmetry is broken to $Z_2$, the three-junction is stable for
$\cos(\theta/2)=t_n/2t_n'$, where $\theta$ is the angle between the modified
defects, with tensions changed from $t_n$ to $t_n'$. This implies that
$t_n'>t_n$ for $\theta>2\pi/3$, and $t_n'\leq t_n$ for $\theta\leq2\pi/3$.
If the vacuum state deviates significantly from its $Z_3$-symmetric position,
we cannot neglect the correction to the energy of the defects, and this would
changes the equilateral hexagons of Fig.~1 to non-equilateral hexagons.
However, if the vacuum state deviates slightly from its $Z_3$-symmetric
position, one may neglect the correction to the energy of the defects.
In this case we are slightly breaking $Z_3\to Z_2$.

We now concentrate on breaking the $Z_2$ symmetry of the host domain wall.
We can do this with the inclusion in the potential of a term odd in $\sigma$,
that slightly removes the degeneracy of the two minima $\sigma=\pm 3/2$.
Thus, the host domain wall bends trying to involve the local minimum, the
false vacuum. To stabilize the non-topological structure we include charged
fields into the system. The way one couples the charged fields is not unique,
but if we choose to add fermions, we can couple them to the $\sigma$ field
in a way such that the projection with $(\phi,\chi)\to(0,0)$ may leave the
model supersymmetric. This is obtained with the superpotential $W(\sigma)$,
with the Yukawa coupling $d^2W/d\sigma^2=(4/3)\sqrt{3}\,\sigma$. In this case
massless fermions bind \cite{jre76} to the host domain wall, and contribute to
stabilize \cite{mba96} the non-topological defect that emerges with the
breaking of the $Z_2$ symmetry.

The breaking of the $Z_2$ symmetry can be done breaking or not the remaining
$Z_3$ symmetry of the model. We examine these two possibilities supposing
that the host domain wall bends under the assumption of spherical symmetry,
becoming a non-topological defect with the standard spherical shape.
This is the minimal surface of genus zero, and according to the Euler
theorem we can only tile the spherical surface with three-junctions
as a regular polygonal network in the three different ways: with 4 triangles,
or 6 squares, or yet 12 pentagons. These three cases preserve the $Z_3$
symmetry of the original network, locally, at the three-junction points.
However, if locally one slightly breaks the $Z_3$ symmetry of the
network to the $Z_2$ one, the three-junctions can now tile the spherical
surface with 12 pentagons and 20 hexagons. We think of breaking the $Z_3$
symmetry minimally, to the $Z_2$ symmetry, through the same mechanism
presented in Fig.~2. Thus, if the symmetry is broken slightly we can
consider the defect tensions as in the regular hexagonal network.

The tiling with 12 pentagons and 20 hexagons generates a spherical structure
that resembles the fullerene, the buckyball composed of sixty carbon atoms.
We visualize the symmetries involved in the spherical structures thinking
of the corresponding dual lattices, which are triangular lattices, but in
the three first cases the triangles are equilateral, while in the fourth case
they are isosceles. We recall that regular heptagons introduce negative
curvature, so they cannot appear when the genus zero surface is minimal.
However, they may for instance spring to generate higher energy states from
the fullerene-like structure, locally roughening the otherwise smooth
spherical surface.

\begin{figure}
\centerline{\psfig{figure=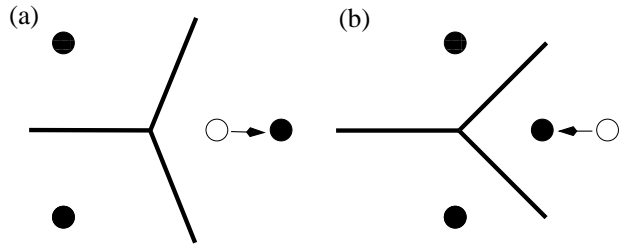,height=3.3cm}}
\vspace{0.4cm}
\caption{The vacuum states (black dots) and the junction that
forms the nested network. (a) and (b) illustrate the only two ways
of breaking the $Z_3\to Z_2$ symmetry.}
\end{figure}

We write the energy of the non-topological structure as
$E^n_{nt}=E^s_{nt}+E_n$, where $E^s_{nt}$ stands for the energy of the
standard non-topological defect, and $E_n$ is the portion due to the nested
network.  We use $E^s_{nt}=E_q+E_h$, which shows the contributions of
the charged fields and of the host domain wall, respectively. We have
$E_h=S\,t_h$ and $E_n=N\,d\,t_{n}$, where $S$ is the area of the spherical
surface, and $N$ and $d$ are the number and length of segments in the nested
network. In the thin wall  approximation the radius $R$ of the non-topological
structure should obey $R\gg l_h$, to make such structure much larger than the
characteristic width of the host domain wall. We introduce the ratio
\be
\frac{E^n_{nt}}{E^s_{nt}}=\left(1+\frac{N}{1+r_{q}}\right)\,\frac{t_n}{t_h}\,
\frac{d}{S}
\ee
with $r_{q}=E_q/E_h$. The non-topological structure nests a network of
defects, which modifies the scenario one gets with the standard domain wall.
The modification depends on the way one couples charged bosons
and fermions to the $\sigma,\phi$, and $\chi$ fields. However, if the
$Z_3$ symmetry is locally broken to the $Z_2$ one, the most probable defect
corresponds to the fullerene or buckyball structure. But if the $Z_3$
symmetry is locally effective, there may be three equilateral structures,
the most probable arising as follows. We consider the simpler case
of plane polygonal structures, identifying the tetrahedron ($i=3$),
cube ($i=4$), and dodecahedron ($i=5$). We introduce $R_{ij}$ as the energy
ratio for the $i$ and $j$ structures. We get
\be
R_{ij}=\frac{1+r_q+\frac{9\sqrt{2}}{16h_i}}{1+r_q+
\frac{9\sqrt{2}}{16h_j}},\;\;\;\;\;
i,j=3,4,5
\ee
Here $h_3,h_4$, and $h_5$ stand for the radius of the
{\it incircle} of the triangle, square, and pentagon, respectively. Energy
favors the tetrahedron, which is self-dual because the network and its dual
are the very same triangular lattice. There are two other configurations,
the octahedron, dual to the cube, and the icosahedron, dual to the
dodecahedron. They do not appear in the $Z_2\times Z_3$ model because
they require four- and five-junctions, respectively.

The present work can be extended in several directions. For instance,
in the $Z_2\times Z_3$ model, if the host domain wall bends cylindrically,
one may get to nanotube-like configurations \cite{sdd99}. Moreover,
we could consider other models, presenting the $Z_2\times Z_k$ symmetry
$(k=4,5,6)$, leading to $k$-junctions. This would allow to tile the plane
with squares $(k=4)$, or triangles $(k=6)$, and the spherical surface with
triangles, as the octahedron $(k=4)$ or the icosahedron $(k=5)$. These
investigations show a new way of modifying the standard domain wall, and this
may find direct application in condensed matter \cite{esc81,sle98,wal97} and
in cosmology \cite{ktu90,vsh96}. In cosmology, for instance, the internal
network modifies the wall energy and changes its cosmological evolution. We
follow Ref.~{\cite{fgg88}} to see that the ratio between the volume pressure
and the surface tension that govern the wall evolution now changes
according to $(p_V/p_T)^n=(p_V/p_T)^s/(1+\alpha\, t_n)$; $\alpha>0$ is a
numerical factor, and $t_n\to0$ recovers the standard case. This shows
that the ending scenario now depends not only on the way the symmetry
is broken, but also on the modification the internal network introduces,
enlarging the possibility of the wall dominating the energy density
before the volume pressure can act. Another line could follow
Ref.~{\cite{lee87}}, asking how fermions could be coupled such
that the corresponding zero modes that inhabit the wall could also bind
to the internal network. Such mechanism would lead to a scenario where the
nested fermions could form a one-dimensional gas inside the nested network,
changing the way the soliton star evolves. We also mention the investigation
concerning pattern formation in cosmology, as in the recent work
{\cite{spa99}}. Our investigation provides other possible scenarios
for pattern formation in the early universe.

We thank C. Furtado, F. Moraes, J. R. S. Nascimento, and R. F. Ribeiro for
discussions, and CAPES, CNPq, and PRONEX for partial support.

\end{document}